\documentclass[apaper,12pt]{article}
\usepackage{caption}
\usepackage{multicol}
\usepackage{graphicx}
\usepackage[left=3.5cm, right=2.5cm, bottom=1.0cm, top=5.0cm]{geometry}
\usepackage{amsmath}
\usepackage{setspace}
\usepackage{array}
\usepackage{subfigure}


\begin{document}
\begin{center}
\noindent {{\Large \bf Finite Size Corrected Relativistic Mean-Field Model and QCD Critical End Point  }}
\vspace{10mm}

{Saeed Uddin{\footnote{$\it saeed\textunderscore jmi@yahoo.co.in$},\, Waseem Bashir}}\\
\vspace{2mm}
{  Department of Physics}\\
{ Jamia Millia Islamia, New Delhi-110025}\\
 India
\end{center}
\vspace{1mm}
\begin{center}
{ Jan Shabir Ahmad}\\
\vspace{2mm}
Department of Physics\\
Baramulla Government College\\
Baramulla, J\&K\\
India
\end{center}
\doublespacing

\begin{abstract}
\noindent We study the effect of finite size of hadrons on the quark hadron phase transition and in particular on the location of the critical end point of such a phase transition. The corrections to the hadronic equation of state are incorporated in a thermodynamic consistent manner for a vander Walls like interaction. For quark gluon plasma phase we take a bag model like  equation of state which takes into account the perturbative interactions among the plasma constituents. We find that for finite sized baryons the first order quark hadron phase transition is not possible for the entire QCD phase diagram. The end point of first order phase coexistence line arises towards the higher chemical potential values in comparison to the point sized baryons, beyond which the transition from hadronic phase to the quark gluon plasma phase might be either crossover or second order phase transition. Our findings are consistent with the finite size scaling ( FSS ) analysis of RHIC data which negates the critical end point with baryon chemical potential values of less than 400 MeV      
\end{abstract}
{\large \bf Introduction}\\

\noindent With the discovery of a possible quark-hadron phase transition under the extreme conditions of temperature and pressure \cite {gross1973}, a numerous efforts have been put forward to map the QCD phase diagram and in particular to locate the critical end point of such a phase transition, where one expects the nature of this transition to change from first to second order or to a smooth crossover \cite {stephanov2005}. However various studies done so far predict different results regarding the location of critical end point in such a deconfining phase transition. This variation stems directly from the treatment of the underlying physics close to the phase transition region involved in these studies, which on a more fundamental level requires a theoretical description in terms of not yet fully understood quantum chromodynamics (QCD) in its non-perturbative regime.

\noindent Keeping this into account an equation of state for a strongly interacting hadronic matter was developed in \cite {saeedwas2012} , where it was argued that for a strongly interacting hadronic matter nearing the point of phase transition, one can in principle ignore the contribution of one pion and one kaon exchanges among various baryons in comparison to the scalar and vector meson exchanges, therefore one could model the strongly interacting hadronic phase with an effective Lagrangian density of the form $\mathcal L= \mathcal {L_{BB}}+ \mathcal {L_{KK}}+ \mathcal {L_{\pi\pi}}$. The EOS for such a strongly interacting hadronic matter was used  to develop the quark-hadron phase transition curve, where it was found that a first order phase transition is possible only upto a certain point, commonly known as critical end point CEP, for such a phase transition, with coordinates coinciding to the CEP as found in one of the variants of lattice gauge theory, LR2 \cite {Fodorkatz2004}

\noindent However it was found that the mesons $\left(\omega,\rho,\phi\right)$ responsible for repulsive interactions among different hadrons were found to be net baryon density $n_B$ dependent, which would imply that for a system with low baryon-chemical potential $\mu_B$ and high temperature values, one can in principle generate large number of particle -antiparticle pairs without any significant repulsive interactions among them \footnote {The effect would be more pronounced for the case of zero baryon-chemical potential, where there are no repulsive interactions among hadrons for any temperature values. }. This is exactly the region where the CEP of quark-hadron phase transition curve is located. Therefore any rectification corresponding to repulsive interactions are expected to effect the location of CEP of such a phase transition. In this work we focus on this problem and construct a  relativistic mean-field ( RMF ) model, which incorporates the finite size volume correction in a thermodynamic consistent manner.

\noindent In this section we present an equation of state ( EOS ), for a strongly interacting hadronic phase and a weakly interacting quark gluon plasma phase used in this work and their respective definition.\\\\

\noindent {\large \bf I.\hspace{3mm} Hadronic Phase  }\\

\noindent To derive the EOS for the hadronic phase consisting of finite sized hadrons we proceed as follows. First of all we take into consideration a system of Boltzmann particles with two particle interactions. The quantum mechanical generalization of such an EOS as applied to the hadronic matter will be written in a form similar to the one derived using Relativistic-Mean Field theory with a correction term corresponding to the finite size of hadrons.\\\\       

\noindent {\large \bf II. \hspace{3mm} Boltzmann Particles with two Particle Interaction}\\

\noindent  Let us consider a system of Boltzmann-particles, with two particle interactions, in thermal and chemical equilibrium at a temperature `T'. A thermodynamic description of this system can be obtained using a grand-canonical partition function, \cite {mayer1977} 

\begin{align*}
\ln Z_G = V\left[n+ \sum_{i=2}^\infty B_i(T)n^i\right]   \tag{1}
\end{align*}\\
\noindent  here `V' is the volume of the system, `n' is the particle number density and `$B_i$(T)' are the viral coefficients. Using a single particle partition function `$z_0$',

\begin{align*}
z_0= g \int \frac{d^3k}{(2\pi)^3} exp\left( \frac{-(k^2 +m^2)^{\frac{1}{2}}-\mu}{T}\right)   \tag{2} 
\end{align*}
\noindent Eq.(1) reduces to,

\begin{align*}
\ln z_0 = \ln (n) + \sum_{i=2}^ \infty \frac{i}{i-1} B_i(T)n^{i-1}  \tag{3}
\end{align*}

\noindent Now if we define the expansions,

\begin{align*}
 P_{cl}(n,T)= T\sum_{i=2} ^\infty B_i(T) n^i,\hspace{10mm}   U_{cl}(n,T)= T\sum_{i=2} ^\infty \frac{i}{i-1} B_i(T)n^{i-1} \tag{4} 
\end{align*}

\noindent the pressure `P' and number density `n'  takes the form \cite {anchsuh1995}

\begin{align*}
P(T,\mu)=& \frac{T}{V}\ln Z_G = Tn(T,\mu) + P_{cl}(n,T)  \tag{5} 
\end{align*}
\begin{align*}
n(T, \mu)=& z_0 \exp \left(-\frac{U_{cl}(n,T)}{T}\right)= g\int \frac{d^3k}{(2\pi)^3}\exp \left(\frac{-(k^2+m^2)^\frac{1}{2}-\mu + U_{cl}(n,T)}{T}\right)  \tag{6}
\end{align*}

\noindent where `g' and `$\mu$' are the degeneracy and chemical potential respectively. Now one can easily verify that the functions, $U_{cl}(n,T)$ and $P_{cl}(n,T)$ are related by \cite {anchsuh1995}

\begin{align*}
{n\frac {\partial U_{cl}(n,T)}{\partial n}= \frac{\partial P_{cl} (n,T)}{\partial n}}  \tag{7}
\end{align*}
{\noindent  Here it is quite clear that with the knowledge of the ${P_{cl}(n,T)}$ and $U_{cl}(n,T)$,  one can in principle calculate the pressure `P' using Eq.(5), however because of Eq.(7), only one of the two functions must be known a priori. For illustration consider a vander-walls EOS,}\\
{\begin{align*}
P= \frac{nT}{1-v_0 n} - \frac{a}{n^2}  \tag{8}
\end{align*}}\\
{ where n, is the particle number density and $v_0$ is the volume of each particle. Equating with Eq.(5) one gets} 

{
\begin{align*}
P_{cl}(n,T)= nT \frac{v_0 n}{1-v_0 n} + \frac{a}{n^2}  \tag{9}
\end{align*}}\\
{ It is clear that the contribution corresponding to the finite volume of particles is contained in first term, denoting it by $P_{v_0}$ and the second term by $P_a$, which takes into account the attractive interactions among the particles, one can calculate the function $U_{cl}(n,T)$ as,}\\
{\begin{align*}
U_{cl}(n,T)= \int \frac{1}{n'} \frac{\partial}{\partial n'} \left(P_{v_0} + P_a \right) dn'
= \int \frac{1}{n'}\frac{\partial P_{v_0}}{\partial n'}  dn' + \int \frac{1}{n'} \frac{\partial P_a}{\partial n'}dn'
= U_{v_0} + U_a         \tag{10}
\end{align*}}\\
{therefore with the knowledge of $P_{v_0}$ one can directly calculate the corresponding function $U_{v_0}$ or vice versa. The generalization of above results to quantum statistics can be readily obtained and are as follows, \cite {anchsuh1995}}

{ \begin{align*}
 P(T,\mu)=& \frac{gT}{a} \int \frac{d^3k}{(2\pi)^3} \ln \left[1+ a\exp\left(\frac{(k^2+m^2)^{\frac{1}{2}} + U_Q - \mu}{T}\right)\right]  +P_Q\\ =& \frac{g}{3} \int \frac{d^3k}{(2\pi)^3} \frac{k^2}{(k^2 +m^2)^{\frac{1}{2}}} f(k) +P_Q  
\tag{11}\\ \\
n(T,\mu)= & g\int \frac{d^3k}{(2\pi)^3} f(k)  \tag{12}
\end{align*}}\\
{and the distribution function is given by ${\small  f(k)= \left\{\exp\left(\frac{(k^2 +m^2)^\frac{1}{2} + U_Q -\mu}{T}\right)+a\right\}^{-1}}$. Here { a}= +1,\,\,-1 for fermions and bosons respectively. The functions ${\tiny {U_Q}}$ and ${\small { P_Q}}$ are the quantum mechanical analogs of the classical functions ${\small U_{cl}(n,T)}$ and ${\small P_{cl}(n,T)}$ respectively. It can be readily verified that for the limit of \,\, ${ \small  {a \to 0}}$, the pressure `P' and number density `n' ( Eq.(11),\, Eq.(12)) approach their classical limit (Eq.(5),\, Eq.(6)) provided one has following set of relations}
\begin{align*}  
{\lim_{a \to 0}} U_Q= U_{cl}(n,T); \hspace{5mm}  {\lim_{a \to 0}} P_Q= P_{cl}(n,T)  \tag{13}
\end{align*}\\
\noindent {\large \bf III. \hspace{3mm} Quantum Statistics of Interacting Hadronic Matter}\\

{\noindent For an interacting hadronic matter, comprising of baryons, pions and Kaons along with their anti-particles, in chemical and thermal equilibrium, we write the total pressure `P' as
\begin{align*}
 P_{QS} = P =  P_{B,\overline B} +P_{b,\overline b} +P_Q   \tag{14}
\end{align*}

{\noindent where, the baryonic contribution to the pressure is}
{\begin{align*}
P_{B,\overline B}=& \sum_B  \frac{gT}{a} \int \frac{d^3k}{(2\pi)^3} \ln \left\{ 1+ a \exp \left(\frac{-(k^2+m_B^2)^\frac{1}{2}-U_Q +\mu_{B}}{T}\right) \right\} \\&+ \frac{gT}{a} \int \frac{d^3k}{(2\pi)^3} \ln \left\{1+ a\exp \left(\frac{-(k^2+m_B^2)^\frac{1}{2}-{\overline U}_Q-\mu_B}{T}\right)\right\}
\end{align*}}

{\noindent and the bosonic contribution is,}

{\begin{align*}
P_{b,\overline b}=& \sum_b  \frac{gT}{a} \int \frac{d^3k}{(2\pi)^3} \ln \left\{ 1+ a \exp \left(\frac{-(k^2+m_b^2)^\frac{1}{2}-U_Q +\mu_{b}}{T}\right) \right\}\\& + \frac{gT}{a} \int \frac{d^3k}{(2\pi)^3}  \times \ln \left\{1+ a\exp \left(\frac{-(k^2+m_b^2)^\frac{1}{2}-{\overline U}_Q-\mu_b}{T}\right)\right\}
\end{align*}}

{\noindent The above expression for the total pressure `P', follows directly from Eq.(11), where the contribution from all particles (hadrons) has been taken into account. Here, ${\bf m_B}$, ${\bf m_b}$ are the masses of the free baryons and bosons carrying a chemical potential,  ${\bf \mu_B}$, \,\, ${\bf \mu_b}$ respectively. For a system of interacting particles and anti-particles one expects these variables to change and take some new effective values. Such a variation in the present model is controlled by the functions, $U_Q$ and ${\overline U}_Q$ respectively. Here, for baryons, we sum up over the entire baryon octet $\left(N,\Lambda, \Sigma,\Xi, \Delta \right)$ and for bosons, we include Kaons,\,\,$ k^+, k^-, k^0, {\overline k}^0$ and pions   $\pi^+, \pi^0, \pi^-$ only.}

{\noindent Now , for a system of interacting hadrons, the function  $ U_Q $, takes the form,}

{\begin{align*}
U_Q \equiv& U_Q \left(v_B,\, v_b,\,T \right)   \tag{15a}\\
= &U_Q \left(n^H,n_B,n_b,n^{S_B},n^{S_b},T \right)    \tag{15b}
\end{align*}}

{\noindent where, $v_B$, $v_b$ are the effective chemical potential's of baryons and the bosons. $n^H$ is the total number density of hadrons, $n_B$ and $n_b$ are the net-baryon density and net-boson density. Also $n^{S_b}, n^{S_B}$ are the scalar density of bosons and baryons respectively.

\noindent For a system of non-interacting particles, as soon as the interactions (repulsive/ attractive) are are turned on, it is clear that the chemical potential of each particle must change, and as the equilibrium state, starts to set in, the effective chemical potential for each particle should approach an equilibrium value. Thus for a system of interacting hadrons, in the equilibrium state, the chemical potential for baryons and bosons must take some equilibrium value, say,  $v_B$ and $v_b$ respectively.  Therefore we define $U_Q$ as a function of effective chemical potentials, $v_B$ and $v_b$, rather than chemical potential corresponding to the free baryons and bosons.
Further keeping into account the functional dependence of the classical function $U_{cl}$, Eq.(4), one can safely assume that the quantum function $U_Q$, with a limiting value of, $\lim_{a\rightarrow 0} U_{Q}= U_{cl} $, must also  be a function of number density as well, in general. Now, for the hadronic system under consideration, which consists of baryons, pions and kaons, it is quite clear that the variable T,  controls the total number density $n^H$ of hadrons in the system , and also the effective chemical potential's, $\,\,  v_B$ and $ v_b$, control the net-baryon density and net-boson density. In addition, these variables can also be thought of to give rise to scalar baryon density $n^{S_B}$ and an analogous quantity for bosons $n^{S_b}$ as well. Therefore it is quite feasible to define $U_Q$ as in  Eq.(15b). This seems to us, the most general form of the field $U_Q$.}

{\noindent Now, we write the function $U_Q$ as, \footnote{Appendix A.1}} 
{\begin{align*}
U_Q= U_1\left( n^H,T \right) + U_2\left(n^{S_B}, n^{S_b}, T \right) + U_3\left(n_B,n_b,n^{S_b},T \right)  \tag{16a}
\end{align*}}
{\noindent with the symmetry property, $U_3 \rightarrow -U_3$, under the transformation, $\mu_B \rightarrow -\mu_B$  or $\mu_b \rightarrow -\mu_b$ \footnote {this follows directly from the symmetry of pressure under this transformation.}\\
{ \noindent Therefore we define,}}
{
\begin{align*}
{\overline U}_Q= U_1\left( n^H,T \right) + U_2\left(n^{S_B}, n^{S_b}, T \right) - U_3\left(n_B,n_b,n^{S_b},T \right). \tag{16b}
\end{align*}}
{\noindent Now for the thermodynamic consistency, it is clear that the thermodynamic relation }

{\begin{align*}
\frac{\partial P}{\partial \mu_B} \Bigg|_{\mu_b, T} = n_B     \tag{17}
\end{align*}}

{\noindent must always hold.}
{\noindent Therefore for pressure `P' as defined in Eq.(14),  with the functions $U_Q$  and  ${\overline U}_Q $ as defined in Eq.(16a) and  Eq.(16b), we  have \footnote{ here we have used,\,\,\, $\frac{\partial P_Q}{\partial \mu_B}= \frac{\partial P_Q}{\partial n^H} \frac{\partial n^H}{\partial \mu_B} + \frac{\partial P_Q}{\partial n^{S_B}} \frac{\partial n^{S_B}}{\partial \mu_B}+ \frac{\partial P_Q}{\partial n_B}\frac{\partial n_B}{\partial \mu_B}$\,\, where, ${\tiny P_Q=P_Q (n^H,\,n^{S_B},\, n^{S_b},\, n_b,T)}$} }

{
\begin{align*}
\frac{\partial P}{\partial \mu_B} \Bigg|_{\mu_b,\,T}= & \,\, n_B + \left( \frac{\partial P_Q}{\partial n^H}- n^H \frac{\partial U_1}{\partial n^H}\right)\frac{\partial n^H}{\partial \mu_B}+  \left(\frac{\partial P_Q}{\partial n^{S_B}}-n^H \frac{\partial U_2}{\partial n^{S_B}}\right)\frac{\partial n^{S_B}}{\partial \mu_B} \\& + \left(\frac{\partial P_Q}{\partial n_B}- n_B \frac{\partial U_3}{\partial n_B}-n_b \frac{\partial U_3}{\partial n_B} \right)\frac{\partial n _B}{\partial \mu_B}.  \tag{18}
\end{align*}}

{\noindent Therefore for thermodynamic consistency one can write}

{
\begin{align*}
\left( \frac{\partial P_Q}{\partial n^H}- n^H \frac{\partial U_1}{\partial n^H}\right)\frac{\partial n^H}{\partial \mu_B}+ \left(\frac{\partial P_Q}{\partial n^{S_B}}-n^H \frac{\partial U_2}{\partial n^{S_B}}\right)\frac{\partial n^{S_B}}{\partial \mu_B}  +\left(\frac{\partial P_Q}{\partial n_B}- n_B \frac{\partial U_3}{\partial n_B}-n_b \frac{\partial U_3}{\partial n_B} \right)\frac{\partial n_B}{\partial \mu_B}=0.    
\end{align*}}

{\noindent Now because of the independence of the variables ($n^{H}, n^{S_B}, n_B$) and therefore of the their derivatives with respect to $\mu_B$ as well, the terms in the brackets should vanish separately. Equating first term to zero, we get,}
{\begin{align*}
\frac{\partial P_Q}{\partial n^H} - n^H \frac{\partial U_1}{\partial n^H}= 0 
\end{align*}} 
{\noindent which after integration yields,}
{\begin{align*}
P_Q= P_{Q 1}(n^H, T)+ P_{Q 2} (n_B, n_b, n^{S_b}, n^{S_B}, T)  \tag{19}
\end{align*}}

{\noindent Therefore the total pressure `P' can be written as, }
{\begin{align*}
P_{QS}= P_{B, \overline{B}} + P_{b, \overline{b}} + P_{Q1} +P_{Q2}  \tag{20}
\end{align*}}
{\noindent where, the baryonic contribution $P_{B,\overline B}$ is,}
{\begin{align*}
P_{B, \overline B}= & \sum_B \frac{gT}{a} \int \frac{d^3k}{(2 \pi)^3} \ln \left\{1 + a\exp \left(\frac{-E^*_B-U_1+v_B}{T}\right)\right\}\\&+ \frac{gT}{a}\int \frac{d^3k}{(2\pi)^3} \ln \left\{ 1+ a\exp \left(\frac{-E^*_B -U_1 -v_B}{T}\right)\right\}\\
=& \sum_B \frac{g}{3} \int \frac{d^3k}{(2\pi)^3} \frac{k^2}{E^*_B}\left(f_B +f_{\overline B}\right) \tag{21}
\end{align*}}
\noindent and for the bosonic contribution $P_{b, \overline b}$, we have,

{
\begin{align*}
P_{b, \overline b}= & \sum_b \frac{gT}{a} \int \frac{d^3k}{(2 \pi)^3} \ln \left\{1 + a\exp \left(\frac{-E^*_b-U_1+v_b}{T}\right)\right\}\\&+\frac{gT}{a}\int \frac{d^3k}{(2\pi)^3} \ln \left\{ 1+ a\exp \left(\frac{-E^*_b -U_1 -v_b}{T}\right)\right\} \\=& \sum_b \frac{g}{3} \int \frac{d^3k}{(2\pi)^3} \frac{k^2}{E^*_b}\left(f_b +f_{\overline b}\right) \tag{22}
\end{align*}}

{\noindent here $E^*_B= E_B + U_2$ and $E^*_b= E_b + U_2 $ are taken to be the, in-medium  effective energies of baryons and bosons. Also $v_B= \mu_B- U_3$ and $v_b= \mu_b- U_3$ represent the effective chemical potential of baryons and bosons respectively.}\\

\noindent {\large \bf IV. \hspace{3mm} RMF Model and the Finite Size Correction}\\
{ \noindent In an RMF model for strongly interacting hadrons ( baryons + pions+ kaons ),the Lagrangian density is
\begin{align*}
\mathcal L^{Total}=\mathcal L_B+\mathcal L_K+\mathcal L_\pi \tag{23}
\end{align*}
\noindent where
\begin{align*}
\mathcal L_B=& \sum_B {{\overline\Psi}_B}[i\gamma_{\mu}\partial^{\mu} - m_B -g_{\sigma B}\sigma+
g_{\sigma^{*} B}\sigma^* -g_{\omega B}\gamma_\mu\omega^\mu-g_{\phi B}\gamma_\mu \phi^\mu -
g_{\rho B}{\gamma_\mu} {\tau_i} {\rho_i}^ \mu] \Psi_B \\&+
\frac{1}{2}\partial_\mu\sigma \partial^\mu\sigma-\frac{1}{2} m_{\sigma}^2 \sigma^2-
\frac{1}{3} g_2 \sigma^3 - \frac{1}{4}g_3\sigma^4 -
\frac{1}{4}W_{\mu\nu}W^{\mu\nu}+\frac{1}{2}m_\omega^2\omega_\mu\omega^\mu\\&+ \frac{1}{4}c_3(\omega_\mu\omega^\mu)^2-
 \frac{1}{4}R_{i\mu\nu}R_i^{\mu\nu}+\frac{1}{2}m_\rho^2\rho_{i\mu}\rho_i^\mu+ \frac{1}{2}\partial_\mu\sigma^* \partial^\mu
\sigma^*-\frac{1}{2}m_{\sigma^*}^2 {\sigma^*}^2\\&-\frac{1}{4}S_{\mu\nu}S^{\mu\nu}+\frac{1}{2}m_\phi^2\phi_\mu\phi^\mu   \tag{24}
\end{align*}
here {$S^{\mu\nu}= \partial^{\mu}\phi^{\nu}-\partial^{\nu}\phi^{\mu}$, $R^{a \mu \nu}=\partial^{\mu}\rho^{a\nu}-\partial^{\nu}\rho^{a\mu}+g_{\rho}\epsilon^{abc}\rho^{b\mu}\rho^{c\nu}$  and $W^{\mu \nu}= \partial^{\mu}\omega^{\nu}-\partial^{\nu}\omega^{\mu}$} { and the sum is carried over the entire baryon octet.\\
Also
\begin{align*}
 \mathcal L_K= &\sum_K \partial_{\mu}{ K^ \dagger} \partial^{\mu}K- m_K^2{ K^ \dagger}K- g_{\sigma K}m_K{ K ^ \dagger}K \sigma -g_{\sigma^*K}m_K{ K ^ \dagger}K\sigma^*-g_{\omega K}{ K ^ \dagger}i \overleftrightarrow{\partial_\mu} K \omega^\mu\\&- g_{\rho K}{K ^ \dagger} {\tau_i}i\overleftrightarrow{\partial_\mu}K {\rho_i}^\mu-g_{\phi K}{ K ^ \dagger}i\overleftrightarrow{\partial_\mu}K\phi^\mu+(g_{\omega K}\omega_\mu+ g_{\rho K}\tau_i \rho_{i \mu}+ g_{\phi K}\phi_\mu)^2  {K^ \dagger} K \tag{25}
\end{align*}
\noindent and
\begin{align*}
\mathcal L_\pi= &\sum_\pi \partial_{\mu}{ \pi ^ \dagger} \partial^{\mu}\pi- m_\pi^2{ \pi ^ \dagger}\pi- g_{\sigma \pi}m_\pi{\pi ^ \dagger}\pi \sigma-g_{\omega \pi}{ \pi ^ \dagger}i \overleftrightarrow{\partial_\mu} \pi \omega^\mu-g_{\rho \pi}{\pi ^ \dagger} {\tau_i}i\overleftrightarrow{\partial_\mu}\pi{\rho_i}^\mu 
\\&+\left( g_{\omega \pi}\omega_{\mu}+ g_{\rho \pi} {\tau_i} \rho_{i \mu}\right)^2 {\pi ^ \dagger} \pi \tag{26}
\end{align*}
\noindent The equation of state EOS turns out to be  \cite {saeedwas2012}\\
{\begin{align*}
P=P_{RMF} =& \frac{1}{3}\sum_B\frac{1}{\pi^2}\int\frac{k^4 dk}{\sqrt{k^2 + {{m^*}_B}^2}}\left(f_B + {\overline f}_B\right) + \frac{1}{3}\sum_{b(K, \pi )}\frac{\gamma}{(2\pi)^3}\int \frac{d^3k}{2\omega_b}\left(f_b +{\overline f}_b\right)\\&+ P(n_B, n_b, n^{S_b}, n^{S_B}, T) \tag{27} 
\end{align*}}
\noindent The first two terms are the kinetic terms while the last term has been written in a generalized manner to show the dependency on baryon and boson densities.
{\normalsize \noindent On comparing with Eq.(20) it is quite clear that in the limit of vanishing hadronic volume, the two expressions become equal, i.e for  $v_0 \rightarrow 0$, $P_{RMFT}= P_{QS}$, therefore one can readily correct the above EOS for finite size effects and the resulting expression becomes,    }\\
{\normalsize
\begin{align*}
P_{RMF} =& \frac{1}{3}\sum_B\frac{1}{\pi^2}\int\frac{k^4 dk}{\sqrt{k^2 + {{m^*}_B}^2}}\left(f_B + {\overline f}_B\right) + \frac{1}{3}\sum_{b(K, \pi )}\frac{\gamma}{(2\pi)^3}\int \frac{d^3k}{2\omega_b}\left(f_b +{\overline f}_b\right)\\&+ P(n_B, n_b, n^{S_b}, n^{S_B}, T)+ P(n^H,T) \tag{28}
\end{align*}} \\ 
{\noindent here the modified distribution functions for baryons $f_{B, \overline B}$ and bosons $f_{b, \overline b}$ are given by,} \\
{\footnotesize
\begin{align*}
f_{B(\overline B)}= &\left\{ \exp \left(\frac{(k^2+{m^*_B}^2)^{\frac{1}{2}}+ U(n^H,T)-\mu_{B(\overline B)}}{T}\right)+a\right\}^{-1} \tag{29a} \\ 
f{b(\overline b)}= &\left\{ \exp \left(\frac{(k^2+{m^*_b}^2)^{\frac{1}{2}}+ U(n^H,T)-\mu_{b(\overline b)}}{T}\right)+a\right\}^{-1}\tag{29b}
\end{align*}}

{\normalsize \noindent with the effective chemical potential, $v_B= \mu_B -g_{\omega B} \omega- g_{\phi B} \phi -g_{\rho B} \tau_{3B} \rho$,  $v_{b}= \mu_b -g_{\omega b} \omega -g_{\phi b} \phi- g_{\rho b} \tau_{3b} \rho $ and the effective in-medium mass, $m^*_B= m_B +g_{\sigma B}\sigma +g_{\sigma^*_B}\sigma^*$,  $m^*_b= \left(m_b^2 +m_b(g_{\sigma b}\sigma+ g_{\sigma^* b} \sigma^*)\right)^{\frac{1}{2}}$ for baryons and bosons respectively. Here $\sigma$,\,\, $\sigma^*$,}\,\,$\omega$,\,\,$\rho$\,\, and $\phi$ are the mean-fields present in the system. The field equations can be deduced from the Lagrangian density Eq.(23) by minimizing the corresponding action and are as follows. For sigma ($\sigma$) field
\begin{align*}
 {{m_\sigma}^2}\sigma +g_2 \sigma^2+ g_3 \sigma^3 =& -\sum_B g_{\sigma B}\frac{\gamma_B}{(2\pi)^3} \int\frac{d^3k}{\sqrt{k^2+{m_B^*}^2}} m_B^*\left(n_B +\overline n_B\right)\\&-
\sum_{b=(K,\pi)}g_{\sigma b}m_b \frac{\gamma_b}{({2\pi})^3}\int \frac{d^3k}{2\omega_b} \left(n_b +\overline n_b\right) \tag{30}
\end{align*}
for omega ($\omega$) field
\vspace{3mm}
\begin{align*}
m_\omega^2\omega + c_3 \omega^3=& \sum_B g_{\omega B} \left[\frac{\gamma_B}{2\pi^3} \int d^3 k \left(n_B -\overline n_B\right)\right]+
\sum_{b=(K,\pi)} 2g_{\omega b}\left[\frac{\gamma_b}{(2\pi)^3}\int \frac{d^3 k}{2\omega_b} \right. \\& \times \left. \left(E_b^+n_b + E_b^- \overline n_b\right)\right]-\sum_{b=(K,\pi)}\left(2\omega g_{\omega b}^2+ 2g_{\omega b}g_{\rho b}\rho\tau_3 + 2g_{\omega b}g_{\Phi b}\Phi \right)\\&   \times \left[\frac{\gamma_b}{(2\pi)^3}\int \frac{d^3 k}{2\omega_b} \left(n_b + \overline n_b\right)\right] \tag{31}
\end{align*}
for rho ($\rho$) field
\vspace{3mm}
\begin{align*}
m_\rho^2 \rho= &\sum_B g_{\rho B}\tau_3 \left[\frac{\gamma_B}{(2\pi)^3}\int d^3 k \left(n_B - \overline n_B\right)\right]+
\sum_{b=(K,\pi)} 2 g_{\rho b}\tau_3 \left[\frac{\gamma_b}{(2\pi)^3}\int \frac{d^3 k}{2\omega_b} \right. \\& \left. \times  \left(E_b^+n_b + E_b^- \overline n_b\right)\right] -\sum_{b=(K,\pi)}\left(2g_{\omega_b}g_{\rho b}\omega \tau_3+ 2 g_{\rho b}^2 \tau_3^2 \rho + 2 g_{\Phi b}g_{\rho b}\tau_3 \Phi\right) \\&
\times \left[\frac{\gamma_b}{(2\pi)^3}\int \frac{d^3 k}{2\omega_b}\left(n_b +\overline n_b\right)\right] \tag{32}
\end{align*}
for sigmastar $(\sigma^*)$ field we have
\vspace{3mm}
\begin{align*}
m_{\sigma*}^2 \sigma ^*=& -\sum_B g_{\sigma^* B}\left[\frac{\gamma_B}{(2\pi)^3}\int d^3 k \frac{m_B^*}{\sqrt{k^2 + {m_B^*}^2}}\left(n_B + \overline n_B\right)\right] -
\sum_{b=(K,\pi)} g_{\sigma ^* b} m_b \\& \times \left[ \frac{\gamma_b}{(2\pi)^3}\int \frac{d^3 k}{2{\omega_b}}\left(n_b +\overline n_b\right)\right] \tag{33}
\end{align*}
and for phi ($\phi$) field we have
\vspace{3mm}
\begin{align*}
m_\Phi^2 \Phi = &\sum_B g_{\Phi B}\left[ \frac{\gamma_B}{(2\pi)^3}\int d^3 k \left(n_B - \overline n_B\right)\right]+
 \sum_{b=(K,\pi)} 2 g_{\Phi b}\left[ \frac{\gamma_b}{(2\pi)^3}\int \frac{d^3 k}{2\omega_b} \right. \\& \left. \times \left({E_b}^+ n_b +{E_b}^- \overline n_b\right)\right]- \sum_{b=(K,\pi)} \left(2g_{\omega b}g_{\Phi b}\omega +2g_{\Phi b}g_{\rho b}\tau_3 \rho + 2\Phi g_{\Phi b}^2 \right) \\& \times \left[ \frac{\gamma_b}{(2\pi)^3}\int\frac{d^3 k}{2\omega_ b}\left(n_b +\overline n_b\right)\right] \tag{34}
\end{align*}
For the evaluation for pressure due to hadrons (pions + kaons + pions) all five coupled field equations are to be solved with two given parameters, temperature T and chemical potential $\mu$. The coupling constant parameters are taken to be the one which have been deduced from low energy experiments. Here we want to make clear that it is not known what are the antiparticle couplings, we are assuming that it is of same magnitude as that of particles.\\\\
    
\noindent {\bf \large V.\hspace{3mm} Quark Gluon Plasma Phase  }\\

\noindent For the weakly interacting QGP phase, we use a Bag-model \cite {chodos1974} inspired EOS and restrict the description to three light quark-flavors (u,d,s) and gluons with the perturbative corrections of the order of $\alpha_s$ \cite {ivanov2005}. The pressure and energy density for this system are given by \cite {satarov2009}.

\begin{align*}
P_{QGP}(T,{{\mu}_f})=& \left(1-\frac{4}{5}{\zeta}\right) \frac{N_{g}}{6{{\pi}^2}}\int_0^{\infty}\frac{k^4dk}{\sqrt{{k^2}+{m_{g}}^2}}f_{g}(k)
+\,(1-{\zeta})\sum_{f=1}^{N_{f}}\frac{N_{c}}{3{\pi}^2}\\& \times \int_0^{\infty}\frac{k^4 dk}{\sqrt{k^2+{m_{f}}^2}} \left[f_{q,f}(k)\,+\,{\overline{f}}_{q,f}(k)\right]\,-B \tag{35}      \\\\   
\varepsilon_{QGP}(T,{\mu}_{f})=& (1-{\zeta})\sum_{f=1}^{N_{f}}\frac{N_{c}}{\pi^2}\int_0^{\infty} k^2 dk \, \sqrt{k^2 + {m_{f}}^2}\left[f_{q,f}(k)+ {\overline{f}}_{q,f}(k)\right] \\&+ \left(1-\frac{4}{5}\zeta \right)\frac{N_g}{2\pi^2} \int_0^{\infty}k^2 dk \sqrt{k^2 +m_g^2}\, f_g(k) +\,B  \tag{36}
\end{align*}

\noindent Here $\zeta = \alpha_s$ is the model parameter and accounts for the strength of interaction present in the system. $N_g= 2(N_c^2-1), N_c$ are the number of transverse gluons and quark colors respectively. $m_f$  is the  quark mas of flavor `f' and $m_g$ is the gluon mass. The distribution functions for the quarks and gluons are $f_{q,f}(k)$ and $f_g(k)$ respectively. Here `B' is the bag constant and is second free parameter in our model. Now because of the energy scales present in the system , it is clear that one can treat two light quarks u and d to be almost mass-less. Therefore solving the above set of equations for two mass-less quarks (u, d) and an s-quark of finite mass $m_s$, the expressions for pressure and energy density can be written as,
\begin{align*} 
P(T,{\mu}) =& \bar{N_{g}}\frac{\pi^2 T^4}{90}+ \bar{N_{f}}\left(\frac{7}{60}\pi^2 T^4 
           + \frac{1}{2}\mu^2 T^2 + \frac{1}{4{\pi^2}}\mu^4\right)
           + \frac{1-{\zeta}}{\pi^2}\int_{m_{s}}^{\infty}dE \left(E^2 -m_{s}\right)^{\frac{3}{2}} \\& \times \left(f_{k}+{\bar{f}}_{k}\right)- B \tag{37}
\end{align*}
\begin{align*}
\varepsilon({T,{\mu}})= & 3\left(\frac{\bar{N_{g}}\pi^2 T^4}{90}\right) + 3\bar{N_{f}}\left(\frac{7}{60}\pi^2 T^4 + \frac{1}{2} \mu^2 T^2 + \frac{1}{4}\frac{{\mu}^4}{\pi^2}\right) + 3\left(\frac{1-{\zeta}}{\pi^2}\right)\int_{m_s}^{\infty} dE \\& \times E^2 (E^2 - m_{s}^2)^{\frac{1}{2}}  \times \left(f_{k}+\bar{f}_{k}\right) +B \tag{38}
\end{align*}\\
\noindent Here $\bar N_g= 16 \left(1-\frac{4}{5}\zeta \right)$ and $\bar N_f= 2\left(1-\zeta \right)$ are the effective number of gluons and the effective number of light flavors. The third term gives the contribution of strange quark and strange anti-quark respectively. It is worth mentioning here that in this calculation we have taken two light quarks u and d to be mass-less and only strange quark to be of finite mass.\\\\

\noindent {\large \bf VI. \hspace{3mm} Results and Discussions}\\

\noindent To have a theoretical prediction about the QCD phase diagram we will now use the equation of state for the hadronic phase, corrected systematically to take into account the finite size of hadrons as explained above, together with the equation of state for the quark gluon plasma QGP. For hadronic system all five coupled field equations for the $\omega, \sigma, \sigma ^*, \rho $ and $\phi$ field are to be solved in a self consistent manner, for two given parameters , temperature T and baryon chemical potential $\mu$. For a given pair of (T, $\mu$) values a complete set of field values is required to be obtained. Then using these field values different observable's of the hadronic system can be readily computed. In case of QGP phase the problem of obtaining the pressure and energy density is relatively simple and can be achieved quite readily. \\ First of all we present some of the features of QGP phase used in this model. It is clear from the QGP EOS, that there are two model parameters, coupling constant $\zeta= \alpha_s$ and Bag-value `B', which must be determined uniquely to make any sensible prediction using this model. We fix these parameters by referring to some of the earlier work done in this direction. Following  Ref. \cite {saeedwas2012, satarov2009} , we take the value of parameter zeta to be, $\zeta =0.2$, as it was shown that it is only for this value of zeta that the above defined QGP equation of state  shows a behavior consistent with the EOS derived using lattice QCD. Similarly for the parameter B we choose a value of B=344 $MeV/fm^3$ following Ref. \cite {satarov2009}, where it was used successfully for the fluid-dynamical calculation of heavy-ion collisions. Here as already mentioned we will be treating light
 \noindent quark flavors (u , d) as mass-less and the  quark `s' to be of finite mass.  Using these parameters we first of all plot variation of $P_{qgp}$ with interaction parameter $\zeta$ as shown in Fig. 1.

\begin{figure}[h]
\begin{center}
\resizebox{110mm}{75mm}{\includegraphics{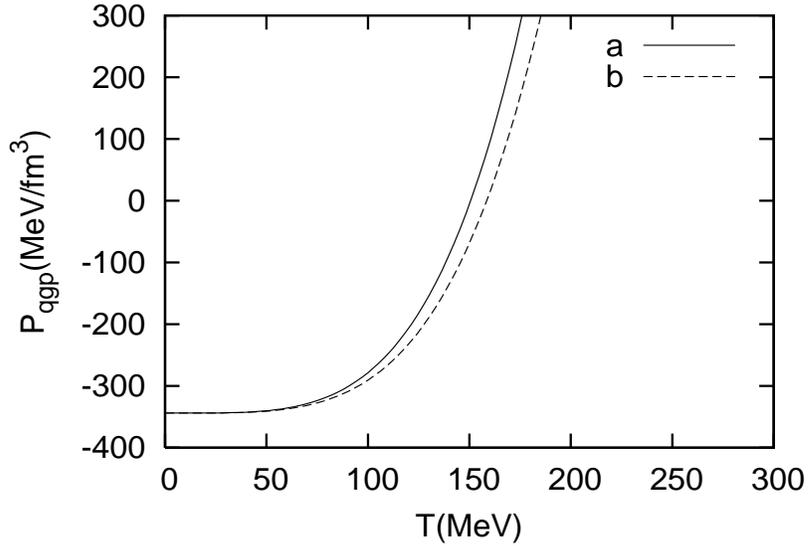}}
\end{center}
\caption{{$P_{QGP}$ v/s T, here $\zeta= 0.0,\,0.2$ for a and b  respectively. Here $\mu_q$=0 MeV.} }
\end{figure}

\noindent  Here we show the variation of pressure in QGP phase for two different cases, a completely non-interacting QGP phase ($\zeta=0.0$) and a QGP phase with perturbative interactions ($\zeta =0.2$). As expected the pressure for the QGP system drops as the interactions are switched on. Starting from a negative value the pressure for the QGP phase becomes positive only after some finite temperature `T', which is the general behavior of bag model equation of state. It is in this region the vacuum pressure is larger than the pressure exerted due to quarks and gluons, which results in the instability of QGP. It is only when $P_{qgp} > 0$ that the QGP phase is in a stable state.

\noindent In Fig. 2 we show the variation of $P_{qgp}$ with temperature `T' for different values of the quark-chemical potential $\mu_q$. Here we have kept the interaction parameter fixed at $\zeta=0.2$. It is this value of the parameter $\zeta$ that we will use throughout our calculation.

\begin{figure}[h]
\begin{center}
\resizebox{110mm}{75mm}{\includegraphics{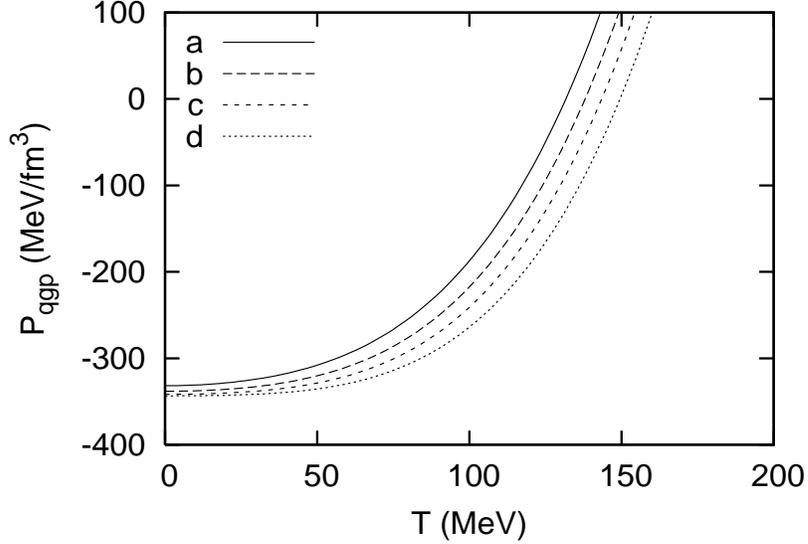}}
\end{center}
\caption{{$P_{QGP}$ v/s T,  $\zeta$ = 0.2 and $\mu_q$= 200, 170, 140, 120 MeV for a,b,c and d  respectively.} }
\end{figure}

\noindent Next, we present some of the features of RMF equation of state. The values of various coupling constants used here are listed in the end. First of all we fix the function $U(n^H,T)$. From Eq.(10) and Eq.(9) we can write, ( after neglecting the vander-walls type attractive interactions) \footnote{In view of the strength of strong interactions one can easily neglect the contribution from such terms.}
 
\begin{align*}
U(n^H,T)= T \frac{v_0 n^H}{1- v_0 n^H}- T\ln (1- v_0 n^H)  \tag{39}
\end{align*}}
{\noindent where the corresponding function $P(n^H, T)$ is, }
{\begin{align*}
P(n^H, T)= n^H T \frac{v_0 n^H}{1-v_0 n^H}   \tag{40}
\end{align*}}

\noindent Here $n^H$ is the total number density of hadrons and $v_0$ is the volume of each hadron. It is for simplicity that we have taken all hadrons to be of equal size. 

\noindent In Fig. 3  we show the variation of $U(n^H,T)$ with temperature for different values of baryon-chemical potential $\mu_B$.\\ 

\begin{figure}[h]
\begin{center}
\resizebox{110mm}{75mm}{\includegraphics{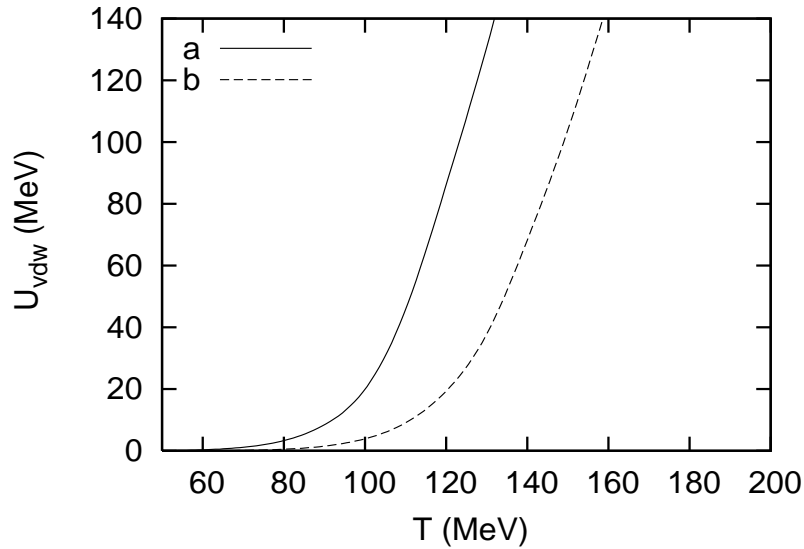}}
\end{center}
\caption{{U$_{vdw}$ v/s T, here $\mu_q= 190, 140 $ MeV for a and b  respectively. } }
\end{figure}

\noindent  Here we have taken only baryons to be of some finite volume $v_0$ and for bosons we choose $v_0= 0$. This is consistent with pauli's exclusion principle from which it follows directly that it is possible for two or more bosons to occupy same position at the same time.

\noindent In Fig. 4  we plot the variation of corresponding mean-field $P(n^H,T)$ with temperature for different values of chemical potential. Among the possible values of baryon radius $r_B$, which lie in the range, $r_B= 0.5-0.8 $ fm \cite {saeed1999}, we have taken  baryon radius to be 0.8 fm and will use same value throughout our calculation.

\begin{figure}[h]
\begin{center}
\resizebox{110mm}{75mm}{\includegraphics{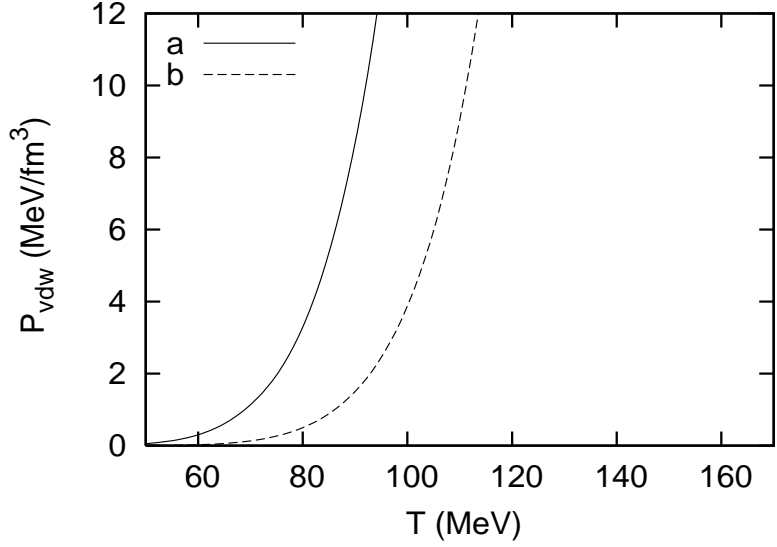}}
\end{center}
\caption{ {P$_{vdw}$ v/s T, here $\mu_q = 190, 140$ MeV for a and b respectively.} }
\end{figure}

\noindent  It is quite evident that the functions $U_{vdw}$ and $P_{vdw}$ have a significant value at higher temperatures, therefore one should expect that the finite size correction should play an important role in the description of quark-hadron phase diagram. Next in Fig. 5 we show the variation of pressure in the hadronic phase with temperature for a fixed value of baryon-chemical potential $\mu_B$.

\begin{figure}[h]
\begin{center}
\resizebox{110mm}{75mm}{\includegraphics{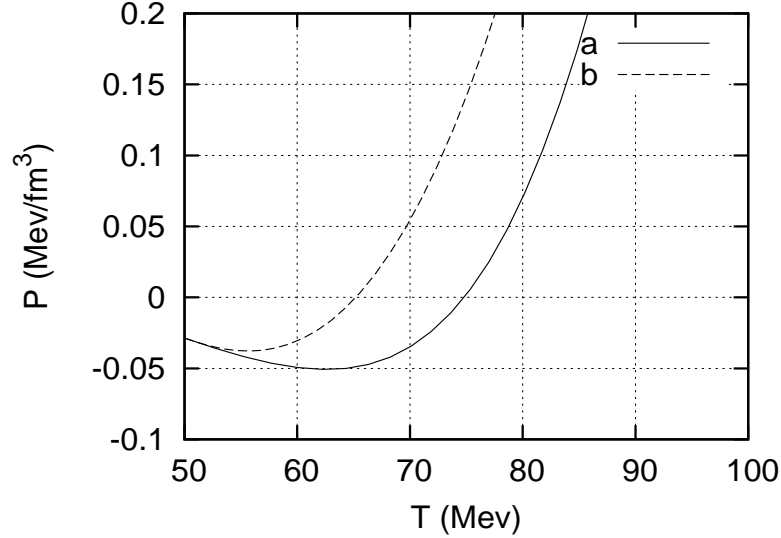}}
\end{center}
{\caption{{ P=P$_{RMFT}$ v/s T, for $\mu_q$ =170 MeV. here hard core radius is $r_0= 0.8fm,\,\, 0fm$ for `a' and `b' respectively.} }}
\end{figure}

\noindent In this figure we have compared the pressure for two different cases. In one we have considered baryons as point particles while in second we have taken baryons to be of finite volume $v_0$. The drop in the pressure for a system of finite sized  baryons  can be attributed to the fact that for any non-negative finite value of mean-field $U (n^H,T)$ the ratio of distribution functions, $\frac{f_B|_{v_0}}{f_B|_{v_0=0}}=e^{-\beta U} $ and  $\frac{f_b|_{v_0}}{f_b|_{v_0=0}}= e^{-\beta U} $ are less than one, which directly imply that, $\frac {P_{RMFT}|_{v_0}}  { P_{RMFT}|_{v_0=0}} < 1$. In the similar fashion one can also show that the P-T curves for other values of baryon-chemical potential as well.

\noindent Now to develop the quark-hadron phase transition diagram we use the Gibbs-criteria of the phase-transition , which requires following set of relations to be valid at the phase equilibrium points, \cite {landaubook}
{\begin{align*}
P_Q= P_H, \hspace{15mm}\mu_Q=\mu_H, \hspace{15mm}T_Q= T_H
\end{align*}}
{\noindent In Fig. 6 and Fig. 7 we show some of the phase co-existence points. The hadronic phase has been amputated beyond pt. B as the entropy density of hadronic matter for this region is negative and  is hence unstable. The stability might arise due to the breaking of this entire hadronic phase into smaller components with pressure P $>$ 0. As the temperature is increased further the system enters in the state with negative total pressure but with positive entropy density. This state is the liquid phase of hadronic matter which is appearing at very high temperatures. This observation for hadronic matter has been also reported in ref \cite {imeada, stock}. For the baryon-chemical potential values less than 420 MeV the point of intersection starts to appear below the $P_{RMFT}=0$ axis. As for this region the pressure due to quarks and gluons is still less than the bag pressure `B', the QGP phase does not exist. An increase of temperature by infinitesimal amount leads to the formation of QGP phase as can be seen from the large slope of P-T curves for this phase. The transition from hadronic to QGP phase therefore occurs without the appearance of the phase coexistence region and hence is a crossover.   }
\begin{figure}[h]
\begin{center}
\resizebox{110mm}{75mm}{\includegraphics{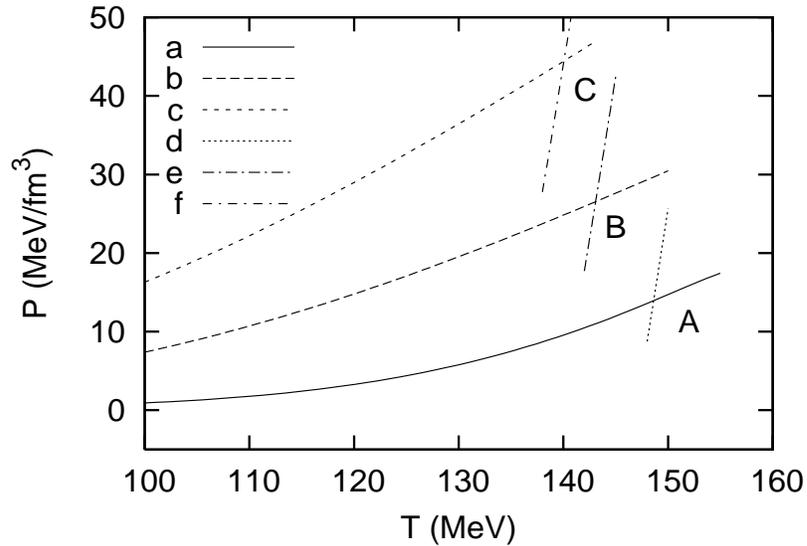}}
\end{center}
{ \caption{{ $P_{RMFT}, P_{QGP}$ intersection plots. A, B, C are the intersection points. Here $\mu_q$= 170, 190, 220 MeV for a,b and c respectively. Here d,e and f  are the $P_{QGP}$ v/s T  curves for the same set of baryon chemical potential values. }}}
\end{figure}
\begin{figure}[h]
\begin{center}
\resizebox{110mm}{75mm}{\includegraphics{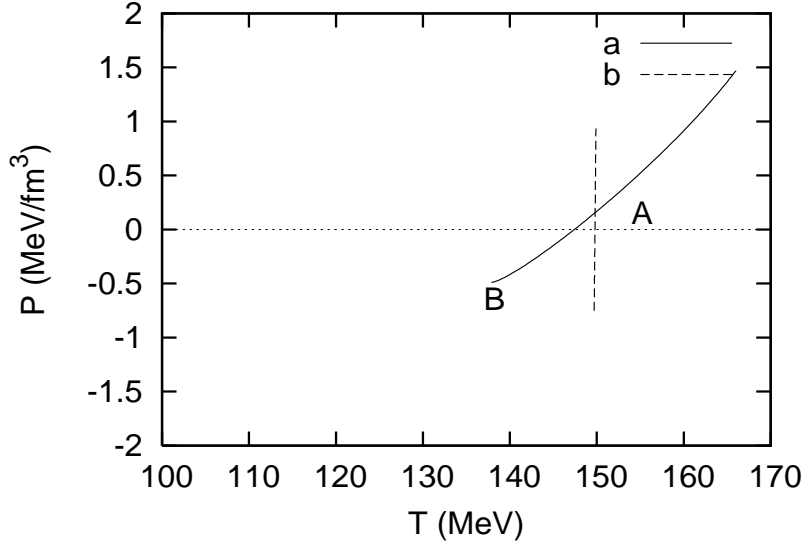}}
\end{center}
{ \caption{ {$P_{RMFT},  P_{QGP} $ for a chemical potential $\mu_B=$140 MeV, here given by a and b respectively. `A' is the point of intersection which is just above the P=0 axis. Here `B' represents the point below which the entropy for the hadronic phase is negative and is therefore amputated  }}}
\end{figure}

\noindent In Fig. 8  we show the quark-hadron phase transition diagram. Here we also compare our result with the one derived for point like hadrons. It is quite clear that the effect of the finite size of baryons is to shift the CEP of the quark-hadron phase transition to higher chemical potential values. Also for any baryon-chemical potential value the temperature values corresponding to the phase coexistence are seen to drop as compared to that for the baryons with no finite size. However one can see the drop is not very large, this fact can be directly attributed to the large slope of P-T curves for the QGP phase. This makes sure that no matter how large the deviation of P-T curves as calculated with RMFTWFS from RMFT is, the intersection  point in both cases lies very much close together.
\begin{figure}[h]
\begin{center}
\resizebox{110mm}{75mm}{\includegraphics{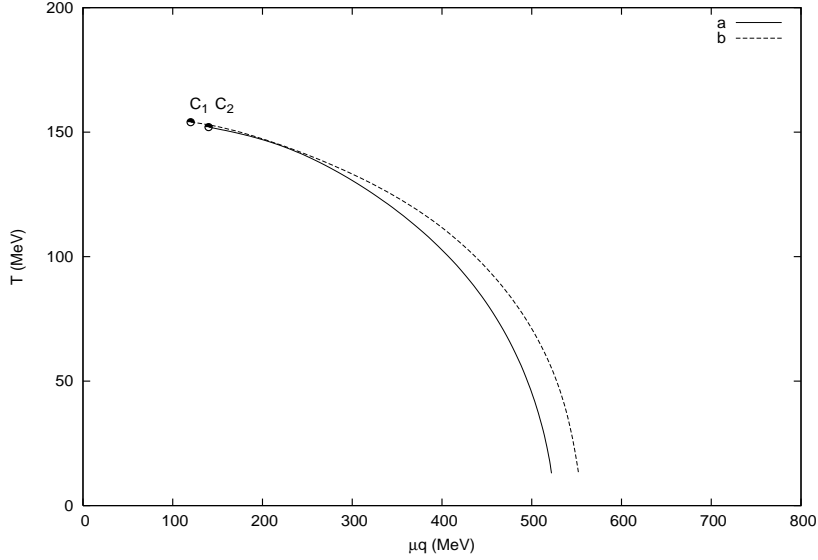}}
\end{center}
{\caption{ {Quark-hadron phase-diagram, here $C_1$ and $C_2$ are the CEP for hard core baryon-radius of $r_B$=0fm and $r_B$=0.8fm , with co-ordinates $C_1$= 120, 154 and $C_2$= 140, 152 respectively. }}}
\end{figure}
{ \noindent  Now in Table.1, we list the results of CEP for the quark-hadron phase transition as obtained in different studies and compare them with our findings here. Keeping into account the versatility of finite-size scaling ( FSS ) \cite {fisher1972, fisherbarber, cardy} and that of Re-normalization Group ( RG ) \cite {wilson1975} methods in general \footnote {The former can be derived quite naturally by applying RG techniques to the critical phenomena \cite {ZinnJustin1985, cardy1966}} which allows one to have the information about the criticality of a system based solely on a very general characteristics and without any detailed description, it is of vital importance to compare any model based results with the recently performed FSS analysis of RHIC data \cite {fraga2011}. In this analysis it was found that the CEP of the quark-hadron phase transition must occur only for baryon chemical potential values of $\mu_B > 400$ MeV.  On comparing with our result it is clear that our result for the critical end point $C_2$ lies well above the lower bound provided by FSS analysis. This is rather strange in view of the fact that our result $C_2$ has been derived in thermodynamic limit of infinite volume and FSS analysis of RHIC data has been performed on a system of finite size. This seems to us the direct consequence of interactions present in the model.
\begin{center}
{\begin{table}[h]
\caption{{ List of CEP's}}
\centering
\begin{tabular}{c  c  c  c}
\hline\hline
Method & $\mu_c$ & $\sqrt {s_{NN}}$ &  $T_C$ \\ [0.5ex]
\hline

FSS  &   $ \sim 400$ MeV & $\sim 5.75$ GeV & \\
$C_1 $ &380 MeV& & 154 MeV  \\
$C_2 $ &420 MeV& & 152 MeV \\
HRG & 156 MeV& &  160 MeV\\
QPM-I & 166 MeV& &183 MeV  \\
QPM-II& 155 MeV& & 166 MeV \\ [1ex]
\hline
\end{tabular}
\label{table:nonlin}
\end{table}}
\end{center}
\noindent One can further clarify the above statement as follows, we know that the size of the system comes into picture while calculating the functions of the form, $\sum f(p,E,m)\rightarrow \frac{1}{h^3} V \int d^3p f(p,E,m)$, which for the interacting system, under the mean-field approximation takes the form, $\frac{1}{h^3} \int d^3 r \int d^3 p f(p,E^*,m^*)$.  Taylor expanding the function `f' we have, $V/h^3\int d^3 p [f(p,E,m) -\alpha \partial{f}/\partial E -\beta \partial f/ \partial m - ... ] $, which can be written as, $1/h^3 V (1- \alpha \partial/ \partial E -\beta \partial/\partial m - ...) \int d^3p f(p,E,m)= 1/h^3 V_{Int} \int d^3p f(p,E,m)$. Clearly $V_{Int}/V < 1$. Therefore for a certain specified class of functions f(p,E,m), which have well defined n-th derivative and are such that $\alpha \partial f/ \partial E, \beta \partial f/\partial m$ are positive dimensionless quantities, a system of interacting particles can be replaced  by a non-interacting system of reduced volume.

\noindent Now we further compare our results with the analysis done in ref. \cite {cpsingh2009}. Here HRG EOS was used for the hadronic phase and bag-model EOS for the QGP phase. It is clear that the CEP obtained in this model is  much below the lower bound provided by the FSS analysis. A similar conclusion can be drawn from the analysis as done in ref. \cite {cpsingh2010} where HRG EOS was again used for the hadronic phase and quasi-particle model QPM-I and QPM-II  \cite {GY, Bannur2006a, Bannur2006b, Golsatz1993, peshkam1996}, were used to study the QGP phase. Here again the CEP obtained is much below the lower bound provided by the FSS analysis.

\noindent Therefore from the above comparisons it becomes clear that the effect of incorporating finite size correction for baryons in the RMFT equation of state has a profound effect on the location of critical end point of the quark-hadron phase transition. The addition of such effects makes sure that the CEP is well within the bound provided by the FSS analysis. More precisely well above the lower bound of FSS analysis. 

\noindent  Now here it is important to mention that the calculations here have been kept up to mean-field level only. This is primarily because of the fact that the couplings are large in our model. However this does not negate the fact that the corrections corresponding to calculations beyond mean-field level should play its part in the location of the critical end point of this quark hadron phase transition. As this requires a detailed study we will pick this in future studies.\\

\noindent {\large \bf VII. \hspace{3mm} Summary}\\

\noindent In this work we have studied the effect of hard core volumes of hadrons on the quark hadron phase transition and in particular on the the critical end point CEP, of such a phase transition. The corrections to the hadronic equation of state, for excluded volume effect are incorporated in a thermodynamic consistent manner for vander walls like interaction. We have found that quantum mechanical generalization of the EOS for Boltzmann particles with two particle interactions can be written in a form similar to the one obtained for an RMF model with an extra term corresponding to the finite size correction. We studied the effect of such hard core volumes  of hadrons on the quark hadron phase diagram. We obtain that the effect of incorporating the finite hard core volumes of the hadrons is to shift the CEP of the quark hadron phase transition to the higher chemical potential values. Our findings are consistent with the finite size scaling of the RHIC data which negates the CEP with baryon chemical potential values of less than 400 MeV.\\     

\noindent {\Large \bf   Acknowledgments}\\
\noindent  Waseem Bashir is thankful to the University Grants Commision for providing Project Fellowship, Saeed Uddin is thankful to the University Grants Commision (UGC), New Delhi, for the Major Research Project grant. Jan Shabir Ahmad is greatful to University Grants Commision, New Delhi, for the financial assistance during the period of deputation.

\newpage

\noindent{ \bf  \large Appendix A.I}\\

{\noindent \textmd From the symmetry of the pressure `P' under the transformation, $\mu_B \rightarrow -\mu_B$ or $\mu_b \rightarrow -\mu_b$, it is clear that the field $U_Q$ must have a contribution from a term which is an odd function of $n_B$ and $n_b$. An obvious choice would be the function $U_Q(n_B,n_b,T)$, where $n_B$ and $n_b$ must always appear in product form. However we choose a more general form of the function $U_Q$ as,}

\begin{align*}
U_Q= U_1(n^H,n^{S_B},n^{S_b},T)+ U_2(n^H,n_B,n_b,n^{S_b},n^{S_B},T)  \tag{A.1}\\
{\overline U}_Q=  U_1(n^H,n^{S_B},n^{S_b},T)- U_2(n^H,n_B,n_b,n^{S_b},n^{S_B},T)  \tag{A.2}
\end{align*}

{\noindent Now, taking the derivative of pressure P with respect to variable $n^H$, we get,}

{\begin{align*}
\frac{\partial P}{\partial n^H}= -n^H \frac{\partial U_Q}{\partial n^H} + \frac{\partial P_Q}{\partial n^H} \tag{A.3}
\end{align*}}

{\noindent Also, for the variation of pressure P with variable $n_B$, one has,}

{
\begin{align*}
\frac{\partial P}{\partial n_B}= -n_B \frac{\partial U_Q}{\partial n_B}- n_b \frac{\partial U_Q}{\partial n_B}+ \frac{\partial P_Q}{\partial n_B}  \tag{A.4}
\end{align*}}

{\noindent {Taking a derivative of A.3 with $n_B$ and A.4 with $n^H$ and by using the independence of $n_B,n^H$ and $n_b$, \footnote {Appendix B.I} and subtracting the resulting equations we have,}}
{\begin{align*}
\left(n^H- n_B-n_b\right)\frac{\partial^2 U_Q}{\partial n^H \partial n_B} = 0  \tag{A.5}
\end{align*}}
{ \noindent Now because of the independence of $n^H,n_B,n_b$ the above equation yields,}
{\begin{align*}
\frac{\partial ^2 U_Q}{\partial n^H \partial n_B} = 0  \tag{A.6}
\end{align*}}
{\noindent which after integrating twice yield,}
{\begin{align*}
U_Q= U(n^H,n^{S_b},n^{S_B},T) + U (n_b, n_B, n^{S_b}, n^{S_B}, T) + U(n^{S_b}, n^{S_B}, T)   \tag{A.7}
\end{align*}}
{\noindent Now using again the independence of pair of variables $n^H$, $n^{S_b}; \,\, n^H,n^{S_B}; \,\, n^{S_B}, n_B,$\footnote{B.I} and following the same procedure as above, one obtains,}

{\begin{align*}
U_Q= U_1(n^H,T) + U_2(n^{S_b}, n^{S_B}, T) + U_3(n_b,n^{S_b}, n_B,T)   \tag{A.8}\\
{\overline U}_Q= U_1(n^H,T) + U_2(n^{S_b}, n^{S_B}, T) - U_3(n_b,n^{S_b}, n_B,T)   \tag{A.9}
\end{align*}}\\

\noindent{ \bf  \large Appendix B.I}\\

{\noindent Let us consider three arbitrary functions  $A \equiv A(A_x, A_y), B\equiv B (B_x, B_y) $ and $C \equiv C(C_x, C_y)$ with the transformation property,\,\,  Det$\left(\frac{\partial A}{\partial B}\right)  \neq $ 0 and  Det$\left(\frac{\partial B}{\partial C}\right)  \neq $ 0, then it can be shown that for the transformation, A$ \rightarrow$ C one has  Det$\left(\frac{\partial A}{\partial C}\right)  \neq $ 0. However if either  Det$\left(\frac{\partial A}{\partial B}\right)  = $ 0  or  Det$\left(\frac{\partial B}{\partial C}\right)  = $ 0, then it can be shown that,  Det$\left(\frac{\partial A}{\partial C}\right)  = $  0}.\footnote{It is for the sake of simplicity that we have kept the functions A,B,C confined to two components only. }  
{  Let us consider a transformation A=A(C), and assume that the determinant of Jacobian matrices, for this transformation, satisfies,}
{\begin{align*}
Det \left(\frac{\partial A}{\partial C}\right)=& 0     \tag{B.1}
\end{align*}}
{\begin{align*}
\Rightarrow \hspace{5mm} \frac{\partial A_x}{\partial C_x} \frac{A_y}{\partial C_y} - \frac{\partial A_x}{\partial C_y} \frac{\partial A_y}{\partial C_x} =0          \tag{B.2}
\end{align*}}
{\noindent Now if there exists the transformations, $A= A(B_x,B_y)$ and $B= B(C_x, C_y)$, then one can write,}
{\begin{align*}
\frac{\partial A_x}{\partial C_x} \frac{\partial A_y}{\partial C_y}\,\, =\,\, & \frac{\partial A_x}{\partial B_x} \frac{\partial B_x}{\partial C_x}\frac{\partial A_y}{\partial B_x} \frac{\partial B_x}{\partial C_y} +  \frac{\partial A_x}{\partial B_x} \frac{\partial B_x}{\partial C_x}\frac{\partial A_y}{\partial B_y} \frac{\partial B_y}{\partial C_y}\\ &+  \frac{\partial A_x}{\partial B_y} \frac{\partial B_y}{\partial C_x}\frac{\partial A_y}{\partial B_x} \frac{\partial B_x}{\partial C_y}+ \frac{\partial A_x}{\partial B_y} \frac{\partial B_y}{\partial C_x}\frac{\partial A_y}{\partial B_y} \frac{\partial B_y}{\partial C_y}              \tag{B.3}
\end{align*}}
{\begin{align*}
\frac{\partial A_x}{\partial C_y} \frac{\partial A_y}{\partial C_x} \,\,=\,\, & \frac{\partial A_x}{\partial B_x} \frac{\partial B_x}{\partial C_y}\frac{\partial A_y}{\partial B_x} \frac{\partial B_x}{\partial C_x} +  \frac{\partial A_x}{\partial B_x} \frac{\partial B_x}{\partial C_y}\frac{\partial A_y}{\partial B_y} \frac{\partial B_y}{\partial C_x} +  \frac{\partial A_x}{\partial B_y} \frac{\partial B_y}{\partial C_y}\frac{\partial A_y}{\partial B_x} \frac{\partial B_x}{\partial C_x} \\&+ \frac{\partial A_x}{\partial B_y} \frac{\partial B_y}{\partial C_y}\frac{\partial A_y}{\partial B_y} \frac{\partial B_y}{\partial C_x}    \tag{B.4}
\end{align*}}
{\noindent therefore,}
\begin{align*}
\frac{\partial A_x}{\partial C_x} \frac{\partial A_y}{\partial C_y}- \frac{\partial A_x}{\partial C_y} \frac{\partial A_y}{\partial C_x}\,\,=\,\,& \left(\frac{\partial A_x}{\partial B_x} \frac{\partial A_y}{\partial B_y}- \frac{\partial A_x}{\partial B_y} \frac{\partial A_y}{\partial B_x}\right) \left(\frac{\partial B_x}{\partial C_x} \frac{\partial B_y}{\partial C_y}- \frac{\partial B_x}{\partial C_y} \frac{\partial B_y}{\partial C_y}\right)\\=& Det \left(\frac{\partial A}{\partial B}\right) Det \left(\frac{\partial B}{\partial C}\right)  
\end{align*}

{ \noindent Now Eq. B.1 becomes,}

{\begin{align*}
Det \left(\frac{\partial A}{\partial C} \right)=  Det \left(\frac{\partial A}{\partial B} \right)  Det \left(\frac{\partial B}{\partial C} \right) = 0       \tag{B.5}
\end{align*}}

{\noindent Which imply that either, ${ Det \left(\frac{\partial A}{\partial B}\right)= 0}$ or $Det \left(\frac{\partial B}{\partial C}\right)= 0$, and also quite simply that,  $Det \left(\frac{\partial A}{\partial B}\right)= 0= Det \left(\frac{\partial B}{\partial C}\right)$. Now if instead of expression B.1, one starts with the assumption that for the transformation A=A(C), determinant of Jacobian is non-zero, i.e $Det \left(\frac{\partial A}{\partial C} \right) \neq 0$,  then it is quite clear from B.4 that, $Det \left(\frac{\partial A}{\partial B} \right) \neq $ 0 and $Det \left(\frac{\partial B}{\partial C}\right) \neq$ 0.}

{ \noindent Now since $n^H$, $n_B$ are the independent parameters and $n_B$,\,\,$n^{S_B}$ are independent also, therefore the determinant of the jacobian matrices for the transformations relating these variables must be non-zero, i.e,}

{\begin{align*}
Det \left(\frac{\partial n^H}{\partial n_B} \right) \neq 0, \hspace {3mm} and \hspace{3mm} Det \left(\frac{\partial n_B}{\partial n^{S_B}} \right) \neq 0
\end{align*}}

{\noindent It therefore follows directly from the above discussion that $n^H$ and $n^{S_B}$ are also independent, i.e}

{\begin{align*}
Det \left(\frac{\partial n^H}{\partial n^{S_B}} \right)\neq 0      \tag{B.6}
\end{align*}}

{\noindent Also for the independent set of variables $ (n^H, n_b)$ and $(n_b, n^{S_b})$, it follows from the above discussion that $(n^H, n^{S_b})$ form another independent pair.}\\

\bibliographystyle{unsrt}
\bibliography{RMFT}

\end{document}